\documentclass{article}%
\usepackage{amsfonts}
\usepackage{amsmath}
\usepackage{amssymb}
\usepackage{graphicx}%
\providecommand{\U}[1]{\protect\rule{.1in}{.1in}}

\begin{document}

\title{Entropic Dynamics: The Schr\"{o}dinger equation and its Bohmian
limit\thanks{Presented at MaxEnt 2015, the 35th International Workshop on
Bayesian Inference and Maximum Entropy Methods in Science and Engineering
(July 19--24, 2015, Potsdam NY, USA). }}
\author{Daniel Bartolomeo and Ariel Caticha\\{\small Department of Physics, University at Albany--SUNY, Albany, NY 12222,
USA}}
\date{}
\maketitle

\begin{abstract}
In the Entropic Dynamics (ED) derivation of the Schr\"{o}dinger equation the
physical input is introduced through constraints that are implemented using
Lagrange multipliers. There is one constraint involving a \textquotedblleft
drift\textquotedblright\ potential that correlates the motions of different
particles and is ultimately responsible for entanglement. The purpose of this
work is to deepen our understanding of the corresponding multiplier
$\alpha^{\prime}$. Its main effect is to control the strength of the drift
relative to the fluctuations. We show that ED exhibits a symmetry: models with
different values of $\alpha^{\prime}$ can lead to the same Schr\"{o}dinger
equation; different \textquotedblleft microscopic\textquotedblright\ or
sub-quantum models lead to the same \textquotedblleft
macroscopic\textquotedblright\ or quantum behavior. In the limit of large
$\alpha^{\prime}$ the drift prevails over the fluctuations and the particles
tend to move along the smooth probability flow lines. Thus ED includes the
causal or Bohmian form of quantum mechanics as a special limiting case.

\end{abstract}

\section{Introduction}

Entropic Dynamics (ED) is a framework that allows the formulation of dynamical
theories as applications of entropic methods of inference \cite{Caticha 2012}.
In the application of ED to derive the Schr\"{o}dinger equation for $N$
particles the physical input is introduced through constraints that are
implemented using Lagrange multipliers \cite{Caticha 2010a}-\cite{Caticha
2015}. There is one set of $N$ constraints, one for each particle, that
control the quantum fluctuations. The central role played by the corresponding
multipliers $\alpha_{n}$ ($n=1\ldots N$) is well understood: they serve to
regulate the flow of time, and the differences among the $\alpha_{n}$ are
associated to differences in the mass of the particles. There is another
constraint involving a \textquotedblleft drift\textquotedblright\ potential
that correlates the motions of different particles. The drift potential
contributes to the phase of the wave function; it is ultimately responsible
for such quantum effects as interference and entanglement. The corresponding
multiplier $\alpha^{\prime}$ is not nearly as well understood and the purpose
of this work is to fill this gap.

We begin with a brief overview of ED following the presentation found in
\cite{Caticha 2014b}. Even at this stage it is clear that the role of
$\alpha^{\prime}$ is to control the strength of the drift relative to the
fluctuations. We show that ED exhibits a symmetry: models with different
values of $\alpha^{\prime}$ can lead to the same Schr\"{o}dinger equation or,
to put it differently, different \textquotedblleft
microscopic\textquotedblright\ or sub-quantum models lead to the same
\textquotedblleft macroscopic\textquotedblright\ or quantum behavior. Then we
argue that the single-valuedness of the quantum wave function restricts the
values that $\alpha^{\prime}$ may take. We conclude by showing in the limit of
large $\alpha^{\prime}$ the drift motion prevails over the fluctuations so
that the particles tend to move along the smooth lines of probability flow.
Thus ED includes the causal or Bohmian form of quantum mechanics as a special
limiting case. Finally we show that ED allows the construction of a hybrid
theory --- a dynamics with quantum fluctuations but no quantum potential
\cite{Nawaz Caticha 2011}. The Bohmian limit of this hybrid theory is fully
equivalent to classical mechanics.

\section{Entropic Dynamics}

As discussed in \cite{Caticha 2014b} we consider the ED of $N$ particles
living in a flat Euclidean space $\mathbf{X}$ with metric $\delta_{ab}$. In ED
particles have definite positions $x_{n}^{a}$ and it is their unknown values
that we wish to infer. (The index $n$ $=1\ldots N$ denotes the particle and
$a=1,2,3$ the spatial coordinate.) The position of the system in configuration
space $\mathbf{X}_{N}=\mathbf{X}\times\ldots\times\mathbf{X}$ is denoted
$x^{A}$ where $A=(n,a)$.

The main assumption is that motion is continuous which means that it can be
analyzed as a sequence of short steps. The method of maximum entropy is used
to find the probability $P(x^{\prime}|x)$ that the system will take a short
step from $x^{A}$ to $x^{\prime A}=x^{A}+\Delta x^{A}$.

The information about the motion is introduced through constraints. The fact
that particles move by taking infinitesimally short steps from $x_{n}^{a}$ to
$x_{n}^{\prime a}=x_{n}^{a}+\Delta x_{n}^{a}$ is imposed through $N$
independent constraints,
\begin{equation}
\langle\Delta x_{n}^{a}\Delta x_{n}^{b}\rangle\delta_{ab}=\kappa_{n}%
~,\qquad(n=1\ldots N)~.~ \label{kappa n}%
\end{equation}
where we shall eventually take the limit $\kappa_{n}\rightarrow0$.
Correlations among the particles are imposed through one additional
constraint,
\begin{equation}
\langle\Delta x^{A}\rangle\partial_{A}\phi=\sum\limits_{n=1}^{N}\left\langle
\Delta x_{n}^{a}\right\rangle \frac{\partial\phi}{\partial x_{n}^{a}}%
=\kappa^{\prime}~, \label{kappa prime}%
\end{equation}
where $\phi$ is the drift potential\footnote{Elsewhere, in the context of
particles with spin, we will see that the potential $\phi(x)$ can be given a
natural geometric interpretation as an angular variable. Its integral over any
closed loop is $%
{\displaystyle\oint}
d\phi=2\pi n$ where $n$ is an integer.} and $\partial_{A}=\partial/\partial
x^{A}=\partial/\partial x_{n}^{a}$. $\kappa^{\prime}$ is another small but for
now unspecified position-independent constant. Eq.(\ref{kappa prime}) is a
single constraint; it acts on the $3N$-dimensional configuration space and is
ultimately responsible for such quantum effects as interference and entanglement.

Already at this early stage we see that ED\ exhibits an epistemic symmetry
that at first sight seems trivial: two rational agents in different epistemic
states can be led to exactly the same inference. Indeed, an agent who imposes
(\ref{kappa prime}) with the pair $(\phi,\kappa^{\prime})$ will assign the
same $P(x^{\prime}|x)$ as another agent who uses the pair $(\tilde{\phi
},\tilde{\kappa}^{\prime})=(C\phi,C\kappa^{\prime})$ where $C$ is some
arbitrary constant.

The result of maximizing entropy leads to
\begin{equation}
P(x^{\prime}|x)=\frac{1}{\zeta}\exp[-\sum_{n}(\frac{1}{2}\alpha_{n}\,\Delta
x_{n}^{a}\Delta x_{n}^{b}\delta_{ab}-\alpha^{\prime}\Delta x_{n}^{a}%
\frac{\partial\phi}{\partial x_{n}^{a}})]~, \label{Prob xp/x a}%
\end{equation}
where $\zeta$ is a normalization constant and $\alpha_{n}$ and $\alpha
^{\prime}$ are Lagrange multipliers. In previous work we took advantage of the
symmetry above and rescaled $\alpha^{\prime}\phi\rightarrow\phi$ which amounts
to choosing $C=1/\alpha^{\prime}$. Here we will keep $\alpha^{\prime}$ explicit.

The successive iteration of many infinitesimal steps to produce a finite
change requires the introduction of time. As discussed in \cite{Caticha
2010a}-\cite{Caticha 2015} entropic time is measured by the fluctuations
themselves which leads to
\begin{equation}
\alpha_{n}=\frac{m_{n}}{\eta\Delta t}~,\label{alpha n}%
\end{equation}
where the particle-specific constants $m_{n}$ will be called \textquotedblleft
masses\textquotedblright\ and $\eta$ is a constant that fixes the units of
time relative to those of length and mass. With this choice of $\alpha_{n}$ a
generic displacement can be expressed as an expected drift plus a
fluctuation,
\begin{equation}
\Delta x^{A}=b^{A}\Delta t+\Delta w^{A}~.\label{Delta x}%
\end{equation}
$b^{A}(x)$ is the drift velocity,
\begin{equation}
\langle\Delta x^{A}\rangle=b^{A}\Delta t\quad\text{with}\quad b^{A}=\frac
{\eta\alpha^{\prime}}{m_{n}}\delta^{AB}\partial_{B}\phi=\eta\alpha^{\prime
}m^{AB}\partial_{B}\phi~,\label{drift velocity}%
\end{equation}
where $m^{AB}$ is the inverse of the \textquotedblleft mass\textquotedblright%
\ tensor, $m_{AB}=m_{n}\delta_{AB}$, and the f{}luctuations $\Delta w^{A}$
satisfy,
\begin{equation}
\langle\Delta w^{A}\rangle=0\quad\text{and}\quad\langle\Delta w^{A}\Delta
w^{B}\rangle=\frac{\eta}{m_{n}}\delta^{AB}\Delta t=\eta m^{AB}\Delta
t~.\label{fluc}%
\end{equation}

These equations show that for very short steps, as $\Delta t\rightarrow0$, the
fluctuations are much larger than the drift ($\Delta w^{A}\sim\Delta t^{1/2}$
while $\langle\Delta x^{A}\rangle\sim\Delta t$) and we have a Brownian motion.
They also show that for fixed $\phi$ the effect of the multiplier
$\alpha^{\prime}$ is to enhance or suppress the drift $b^{A}\Delta t$ relative
to the fluctuations $\Delta w^{A}$.

Having introduced a convenient notion of time through (\ref{alpha n}), the
result of accumulating many changes is that the probability distribution
$\rho(x,t)$ in configuration space obeys a Fokker-Planck equation, (See
\emph{e.g.}, \cite{Caticha 2012}),
\begin{equation}
\partial_{t}\rho=-\partial_{A}\left(  \rho v^{A}\right)  ~,\label{FP b}%
\end{equation}
where $v^{A}$ is the velocity of the probability flow in configuration space
or \emph{current velocity},
\begin{equation}
v^{A}=b^{A}+u^{A}\quad\text{and}\quad u^{A}=-\eta m^{AB}\partial_{B}\log
\rho^{1/2}~
\end{equation}
is the \emph{osmotic velocity}. Since both $b^{A}$ and $u^{A}$ are gradients
the current velocity is a gradient too,%
\begin{equation}
v^{A}=m^{AB}\partial_{B}\Phi\quad\text{where}\quad\Phi=\eta\alpha^{\prime}%
\phi-\eta\log\rho^{1/2}~.\label{curr}%
\end{equation}

The dynamics described by the FP equation (\ref{FP b}) is a standard
diffusion. To describe a \textquotedblleft mechanics\textquotedblright\ we
require that the diffusion be \textquotedblleft
non-dissipative\textquotedblright. This is achieved by an appropriate
readjustment or updating of the constraint (\ref{kappa prime}) after each step
$\Delta t$. The net effect is that the drift potential $\phi$, or equivalently
$\Phi$, is promoted to a fully dynamical degree of freedom. The diffusion is
said to be \textquotedblleft non-dissipative\textquotedblright\ when the
actual updating is implemented by imposing that a certain functional
$\tilde{H}[\rho,\Phi]$ be conserved; in order to offset the entropic change
$\rho\rightarrow\rho+\delta\rho$, one requires a change $\Phi\rightarrow
\Phi+\delta\Phi$ such that
\begin{equation}
\tilde{H}[\rho+\delta\rho,\Phi+\delta\Phi]=\tilde{H}[\rho,\Phi]~.
\end{equation}
As shown in \cite{Caticha 2014b} the requirement that $\tilde{H}$ be conserved
for arbitrary choices of $\rho$ and $\Phi$ implies that the coupled evolution
of $\rho$ and $\Phi$ is given by a conjugate pair of Hamilton's equations,
\begin{equation}
\partial_{t}\rho=\frac{\delta\tilde{H}}{\delta\Phi}\quad\text{and}%
\quad\partial_{t}\Phi=-\frac{\delta\tilde{H}}{\delta\rho}~.\label{Hamilton}%
\end{equation}
When the \textquotedblleft ensemble\textquotedblright\ Hamiltonian $\tilde{H}$
is chosen so that the first equation reproduces the FP equation (\ref{FP b}),
the second becomes a Hamilton-Jacobi equation. Arguments from information
geometry \cite{Caticha 2014b} can be invoked to further specify the form of
the functional $\tilde{H}[\rho,\Phi]$. They suggest that the natural choice of
$\tilde{H}$ is%
\begin{equation}
\tilde{H}[\rho,\Phi]=\int dx\,\left[  \frac{1}{2}\rho m^{AB}\partial_{A}%
\Phi\partial_{B}\Phi+\rho V+\xi m^{AB}\frac{1}{\rho}\partial_{A}\rho
\partial_{B}\rho\right]  ~.\label{Hamiltonian}%
\end{equation}
The first term in the integrand is the \textquotedblleft
kinetic\textquotedblright\ term that reproduces (\ref{FP b}). The second term
is the simplest possible non-trivial interaction, an energy term that is
linear in $\rho$ and introduces the standard potential $V(x)$. The third term
is motivated by information geometry and is called the \textquotedblleft
quantum\textquotedblright\ potential. The parameter $\xi=\hbar^{2}/8$\ turns
out to be crucial: it controls the relative contributions of the two
potentials, it defines the value of what we call Planck's constant $\hbar$,
and it sets the scale that separates quantum from classical regimes.

To conclude this brief review of ED we combine $\rho$ and $\Phi$ into a single
complex function,
\begin{equation}
\Psi=\rho^{1/2}\exp(i\Phi/\hbar)~. \label{psi k}%
\end{equation}
The pair of Hamilton's equations (\ref{Hamilton}) can then be written as a
single complex linear equation,%
\begin{equation}
i\hbar\partial_{t}\Psi=-\frac{\hbar^{2}}{2}m^{AB}\partial_{A}\partial_{B}%
\Psi+V\Psi~, \label{sch c}%
\end{equation}
which we recognize as the Schr\"{o}dinger equation.

\section{Quantized circulation}

An important question is whether the Fokker-Planck and Hamilton-Jacobi
equations, eqs.(\ref{Hamilton}), are fully equivalent to the Schr\"{o}dinger
equation. This point was first raised by Wallstrom \cite{Wallstrom 1989} as an
objection to Nelson's stochastic mechanics \cite{Nelson 1985} and concerns the
single- or multi-valuedness of phases and wave functions. Wallstrom's
objection was that stochastic mechanics leads to phases $\Phi$ and wave
functions $\Psi$ that are either both multi-valued or both single-valued. Both
alternatives are unsatisfactory: quantum mechanics forbids multi-valued wave
functions, while single-valued phases can exclude physically relevant states
(\emph{e.g.}, states with non-zero angular momentum).

The requirement that the wave function $\Psi$ be single-valued amounts to
imposing a quantized circulation condition,
\begin{equation}%
{\displaystyle\oint_{\Gamma}}
d\ell^{A}\partial_{A}\frac{\Phi}{\hbar}=2\pi n~, \label{circ a}%
\end{equation}
where $\Gamma$ is any closed loop in configuration space and $n$ is an
integer. On the other hand, we had earlier briefly mentioned that the drift
potential $\phi$ is to be interpreted as an angle in which case, the integral
over the closed loop $\Gamma$ gives
\begin{equation}%
{\displaystyle\oint_{\Gamma}}
d\ell^{A}\partial_{A}\phi=2\pi n^{\prime}~, \label{circ b}%
\end{equation}
where $n^{\prime}$ is an integer associated to the drift potential $\phi$. We
do not discuss this issue in any detail except to note that this is true when
particle spin is incorporated into the theory. Indeed, as shown by Takabayasi,
a similar result holds for the hydrodynamical formulation of spinning
particles \cite{Takabayasi 1983}. But, if eq.(\ref{circ b}) is true, then we
can use eq.(\ref{curr}) to integrate the phase $d\Phi/\hbar$ over a closed
path. Since $\rho$ is single-valued,
\begin{equation}%
{\displaystyle\oint_{\Gamma}}
d\ell^{A}\partial_{A}\log\rho=0~,
\end{equation}
and we obtain
\begin{equation}%
{\displaystyle\oint_{\Gamma}}
d\ell^{A}\partial_{A}\frac{\Phi}{\hbar}d\ell^{A}=\frac{\eta\alpha^{\prime}%
}{\hbar}%
{\displaystyle\oint_{\Gamma}}
d\ell^{A}\partial_{A}\phi=2\pi n^{\prime}\frac{\eta\alpha^{\prime}}{\hbar}~.
\label{circ c}%
\end{equation}
Comparing (\ref{circ a}) and (\ref{circ c}) we conclude that
\begin{equation}
n^{\prime}\frac{\eta\alpha^{\prime}}{\hbar}=n
\end{equation}
is an integer. Since this must simultaneously hold for all loops $\Gamma$
including loops with arbitrary values of $n^{\prime}$ we conclude that
\begin{equation}
\eta\alpha^{\prime}=N\hbar
\end{equation}
where $N$ is an integer. This is precisely the quantization condition that
leads to full equivalence between ED and the Schr\"{o}dinger equation because
it guarantees that wave functions will remain single-valued even for
multi-valued phases.

\section{The effect of $\alpha^{\prime}$}

The dynamics described by (\ref{Hamilton}) and (\ref{Hamiltonian}), or by the
Schr\"{o}dinger equation (\ref{sch c}) is clearly independent of
$\alpha^{\prime}$ and therefore we have a symmetry. As we see in
eqs.(\ref{drift velocity}) and (\ref{fluc}) different choices of
$\alpha^{\prime}$ lead to different Brownian motions at the sub-quantum or
\textquotedblleft microscopic\textquotedblright\ level. However, they all lead
to the same evolution of $\rho$ and $\Phi$ and the same dynamics --- the same
Schr\"{o}dinger equation --- at the quantum or \textquotedblleft
macroscopic\textquotedblright\ level.

\paragraph{The Bohmian limit}

We can directly study the sub-quantum effect of $\alpha^{\prime}$ in
eqs.(\ref{drift velocity}) and (\ref{fluc}). It is, however, more instructive
to rescale $\eta$ and write $\eta=\tilde{\eta}/\alpha^{\prime}$. Under such
rescaling the $\alpha^{\prime}$ dependence has migrated from the drift to the
fluctuations,
\begin{equation}
\langle\Delta x^{A}\rangle=\tilde{\eta}m^{AB}\partial_{B}\phi\,\Delta
t\quad\text{and}\quad\langle\Delta w^{A}\Delta w^{B}\rangle=\frac{\tilde{\eta
}}{\alpha^{\prime}}m^{AB}\Delta t~. \label{fluc b}%
\end{equation}
Increasing $\alpha^{\prime}$ at fixed $\tilde{\eta}$ has the effect of
suppressing the fluctuations while leaving the drift unaffected. {}In the
limit $\alpha^{\prime}\rightarrow\infty$ we expect the fluctuations to be
negligible; the particles will follow smooth trajectories that do not resemble
a Brownian motion at all.

From eq.(\ref{curr}) we have
\begin{equation}
\Phi=\tilde{\eta}\phi-\frac{\tilde{\eta}}{\alpha^{\prime}}\log\rho^{1/2}~,
\end{equation}
so that for large $\alpha^{\prime}$
\begin{equation}
\Phi\rightarrow\tilde{\eta}\phi\quad\text{and}\quad v^{A}\rightarrow b^{A}~.
\end{equation}
Therefore, for $\alpha^{\prime}\rightarrow\infty$ the current and the drift
velocities coincide. Particles follow smooth trajectories that coincide with
the lines of probability flow. This is exactly the kind of motion postulated
by Bohmian mechanics \cite{Bohm 1952}-\cite{Holland 1993}.

We can therefore claim that, at least formally, entropic dynamics includes
Bohmian mechanics as a special limiting case. However, there are important
differences between ED and Bohmian mechanics that need to be emphasized.
First, it is worth pointing out that the limit $\alpha^{\prime}\rightarrow
\infty$ is a tricky one because it is meant to be taken only after we take the
limit $\Delta t\rightarrow0$. This is what allows us to write differential
equations such as (\ref{FP b}). Thus, no matter how large the (fixed) value of
$\alpha^{\prime}$, entropic dynamics remains \textquotedblleft
entropic\textquotedblright. Even for large $\alpha^{\prime}$ the dynamics is
still driven by fluctuations and at sufficiently microscopic scales the
expected motion is Brownian.

Second, and perhaps even more important, there is a major philosophical
difference: Bohmian mechanics attempts to provide an actual description of
reality, a description of the ontology of the universe as it \textquotedblleft
really\textquotedblright\ is and as it \textquotedblleft
really\textquotedblright\ happens. In the Bohmian view the universe consists
of real particles that have definite positions and their trajectories are
guided by something real, the wave function $\Psi$ \cite{Bohm 1952}.

In contrast, ED is a purely epistemic theory. It does not attempt to describe
the world. Its pragmatic goal is less ambitious and also more realistic: to
make the best possible predictions on the basis of very incomplete
information. In ED the particles also have definite positions and its
formalism includes a function $\Phi$ that behaves as a wave. But $\Phi$ is a
tool for reasoning; it is not meant to represent anything real. There is no
implication that the particles move the way they do because they are guided by
a pilot wave or because they are being pushed around by some stochastic force.
In fact ED is silent on the issue of what causative power is responsible for
the peculiar motion of the particles. What the probability $\rho$ and the
phase $\Phi$ are designed to accomplish is to guide our inferences. They guide
our expectations of where to find the particles but they do not exert any
causal influence on the particles themselves.

\paragraph*{A hybrid theory and the classical limit}

Equation (\ref{Hamiltonian}) includes a parameter $\xi$ that regulates the
strength of the quantum potential. Any non-zero value $\xi>0$ yields a fully
\emph{quantum} mechanics, albeit with differing values of $\hbar$. The value
$\xi=0$ leads, however, to a qualitatively different theory. One might suspect
that $\xi=0$ gives classical mechanics but this is not so.\footnote{By
\textquotedblleft classical\textquotedblright\ mechanics we mean a Newtonian
deterministic mechanics. The $\xi=0$ theory could be called a
\textquotedblleft classical indeterministic mechanics\textquotedblright\ but
this is not useful as it would broaden the meaning of the term `classical' to
cover any theory that is `not-quantum'.} According to equations
(\ref{Hamilton}) and (\ref{Hamiltonian}) for $\xi=0$ the probability $\rho$
follows the gradient of $\Phi$,
\begin{equation}
\partial_{t}\rho=\frac{\delta\tilde{H}}{\delta\Phi}=-\partial_{A}\left(  \rho
v^{A}\right)  \quad\text{with}\quad v^{A}=m^{AB}\partial_{B}\Phi~,
\end{equation}
and $\Phi$ evolves according to the classical Hamilton-Jacobi equation,
\begin{equation}
\partial_{t}\Phi=-\frac{\delta\tilde{H}}{\delta\rho}=-\frac{1}{2}%
m^{AB}\partial_{A}\Phi\partial_{B}\Phi-V~. \label{classical HJ}%
\end{equation}
Therefore the probability $\rho$ flows along the classical path. However,
there is no implication that the particles themselves follow the classical
paths. Indeed, at any instant of time the particles undergo the same
fluctuations, eq.(\ref{fluc b}), that we would expect for any non-zero value
of $\xi$.

The $\xi=0$ model resembles classical mechanics in some respects and quantum
mechanics in others; it is a hybrid theory. Just as in quantum mechanics the
particles follow Brownian paths and the dynamics is a non-dissipative
diffusion; they even satisfy an uncertainty principle \cite{Nawaz Caticha
2011}. On the other hand, just as in classical mechanics, the probability
flows according to paths described by the classical Hamilton-Jacobi equation.
One can even combine $\rho$ and $\Phi$ into a single complex function,
$\Psi=\rho^{1/2}\exp(i\Phi/\hbar)$, and write the coupled evolution of $\rho$
and $\Phi$ in terms of a single complex equation that resembles a
Schr\"{o}dinger equation,
\begin{equation}
i\hbar\partial_{t}\Psi_{k}=-\frac{\hbar^{2}}{2}m^{AB}\partial_{A}\partial
_{B}\Psi_{k}+V\Psi_{k}+\frac{\hbar^{2}}{2}m^{AB}\frac{\partial_{A}\partial
_{B}|\Psi_{k}|}{|\Psi_{k}|}\Psi_{k}~. \label{sch b}%
\end{equation}
But this equation is not linear which means that a central feature of quantum
behavior, the superposition principle, has been lost.

For the hybrid theory too we can take the Bohmian limit $\alpha^{\prime
}\rightarrow\infty$. Increasing $\alpha^{\prime}$ at fixed $\tilde{\eta}$ has
the same effect of suppressing the fluctuations so the particles follow smooth
trajectories that coincide with the lines of probability flow. The one
difference is that for $\xi=0$ the lines of probability flow are determined by
the classical Hamilton-Jacobi equation (\ref{classical HJ}), and therefore the
particles follow classical trajectories. We conclude that the Bohmian limit of
the hybrid theory is classical mechanics. In other words, classical mechanics
is related to the hybrid theory in exactly the same way as Bohmian mechanics
is related to entropic dynamics.

\paragraph*{Acknowledgments}

We would like to thank M. Abedi, C. Cafaro, N. Caticha, S. DiFranzo, A.
Giffin, S. Ipek, D.T. Johnson, K. Knuth, S. Nawaz, M. Reginatto, C. Rodriguez,
and K. Vanslette, for many discussions on entropy, inference and quantum mechanics.

\end{document}